\documentstyle[aasms4,psfig,flushrt]{article}

\newcommand\beq{\begin{equation}} 
\newcommand\eeq{\end{equation}}
 
\def\DeltaRA{|\Delta{\rm RA|}}
\def\DeltaDec{|\Delta{\rm DEC|}}

\received{27 October 1999} 
\accepted{31 January 2000} 

\lefthead{Kong et al.} 
\righthead{Age, Metallicity and Reddening of M81} 

\begin{document}

{ \title{Spatially resolved Spectro-photometry of M81: 
\\ Age, Metallicity and Reddening Maps}

\author{ 
Xu Kong\altaffilmark{1,2}, 
Xu Zhou\altaffilmark{1},
Jiansheng Chen\altaffilmark{1}, 
Fuzhen Cheng\altaffilmark{2,1} ,
Zhaoji Jiang\altaffilmark{1}, 
Jin Zhu\altaffilmark{1}, 
Zhongyuan Zheng\altaffilmark{1}, 
Shude Mao\altaffilmark{3,10}, 
Zhaohui Shang\altaffilmark{1,4},
Xiaohui Fan\altaffilmark{1,5},
Yong-Ik Byun\altaffilmark{6,7},
Rui Chen\altaffilmark{1},
Wen-ping Chen\altaffilmark{6},
Licai Deng\altaffilmark{1},
Jeff J. Hester\altaffilmark{8},
Yong Li\altaffilmark{8},
Weipeng Lin\altaffilmark{1},
Hongjun Su\altaffilmark{1},
Wei-hsin Sun\altaffilmark{6},
Wean-shun Tsay\altaffilmark{6},
Rogier A. Windhorst\altaffilmark{8},
Hong Wu\altaffilmark{1},
Xiaoyang Xia\altaffilmark{9,1},
Wen Xu\altaffilmark{1,8},
Suijian Xue\altaffilmark{1},
Haojing Yan\altaffilmark{1,8},
Zheng Zheng\altaffilmark{1},
and
Zhenglong Zou\altaffilmark{1}
}

\altaffiltext{1}{Beijing Astronomical Observatory and Beijing
Astrophysics Center (BAC), National Astronomical Observational Center, 
Chinese Academy of Sciences, Beijing, 100012, P. R. China }

\altaffiltext{2}{Center for Astrophysics, University of Science and
        Technology of China, Hefei, 230026, P. R. China}
\altaffiltext{3}{Max-Planck-Institute for Astrophysics 
         Karl-Schwarzschild-Strasse 1, 85740, Garching, Germany }
\altaffiltext{4}{Department of Astronomy, University of Texas at Austin,
    Austin, TX 78712}
\altaffiltext{5}{Princeton University Observatory, Princeton, New
    Jersey, 08544}
\altaffiltext{6}{Institute of Astronomy, National Central University,
    Chung-Li, Taiwan}
\altaffiltext{7}{Center for Space Astrophysics and Department of
Astronomy,
Yonsei University, Seoul, 120--749, Korea}
\altaffiltext{8}{Department of Physics and Astronomy, Box 871504, Arizona
    State University, Tempe, AZ  85287--1504}
\altaffiltext{9}{Department of Physics, Tianjin Normal University,
China}
\altaffiltext{10}{Univ. of Manchester,Jodrell Bank Observatory 
Macclesfield, Cheshire SK11 9DL, UK}

\authoremail{xkong@mail.ustc.edu.cn}

\begin{abstract}

In this paper, we present a multi-color photometric study of the nearby
spiral galaxy M81, using images obtained with the Beijing Astronomical
Observatory $60/90$ cm Schmidt Telescope in 13 intermediate-band filters
from 3800 to 10000{\AA}.  The observations cover the whole area of M81
with a total integration of 51 hours from February 1995 to February
1997. This provides a multi-color map of M81 in pixels of $1\arcsec.7
\times 1\arcsec.7$.  Using theoretical stellar population synthesis
models, we demonstrate that some BATC colors and color indices can be
used to disentangle the age and metallicity effect.  We compare in detail
the observed properties of M81 with the predictions from population
synthesis models and quantify the relative chemical abundance, age and
reddening distributions for different components of M81.  We find that
the metallicity of M81 is about $Z=0.03$ with no significant difference
over the whole galaxy.  In contrast, an age gradient 
is found between stellar populations of the central regions and of the
bulge and disk regions of M81: the stellar population in its central
regions is older than 8 Gyr while the disk stars are considerably
younger, $\sim 2$ Gyr.  We also give the reddening distribution in M81.
Some dust lanes are found in the galaxy bulge region and the reddening
in the outer disk is higher than that in the central regions.

\end{abstract}

\keywords{galaxies: abundances -- galaxies: evolution -- galaxies:
individual (M81) -- galaxies: stellar  content -- (ISM:) dust, reddening}

\section{INTRODUCTION}

Spatially resolved information about the age, metallicity and interstellar
medium reddening of galaxies is a powerful tool to study galaxy
evolution since it provides essential clues in star formation history,
chemical composition and enrichment history and environment of galaxies
(\cite{Buzzoni89}).  To obtain such information, we need to know the
stellar component and the overall properties of stellar populations
(\cite{Leitherer95}). Ideally one would like to study
resolved individual stars in galaxies. However, given the limited spatial
resolution of current telescopes, this is only possible for a few very
nearby galaxies. As a result, the stellar content of even some relatively
simple galaxies remain to be unraveled (\cite{Thuan91}). Information of
stellar population and star formation history in these galaxies, however,
can still be obtained from studying the {\it integrated} properties of
the stars (e.g. \cite{Schmitt96}, \cite{Goerdt98}).

Since the pioneering work of Tinsley (1972) and Searle et al. (1973),
evolutionary population synthesis has become a standard technique to
study the stellar populations of galaxies. This is a result of improvements
in the theory of the chemical evolution of galaxies, star formation,
stellar evolution and atmospheres, and the development of synthesis
algorithms and the availability of various evolutionary synthesis models.
A comprehensive compilation of such models was published by Leitherer et
al. (1996) and Kennicutt (1998). Widely used models include those from
the Padova and Geneva group (e.g. \cite{Schaerer97}, \cite{Schaerer98},
\cite{Bressan96}, and \cite{Chiosi98}), GISSEL96 (\cite{Charlot91},
\cite{Bruzual93}, \cite{Bruzual96}), PEGASE (\cite{Fioc97}) and
STARBURST99 (\cite{Leitherer99}).

Many previous studies of integrated stellar populations use
spectroscopic data, usually for limited regions in galaxies.  In this
paper, we will, instead, use multi-color photometry to probe the stellar
populations.  The multi-color photometry provides accurate spectral energy
distributions (SEDs) for the {\it whole} galaxy, although at low spectral
resolution. We shall demonstrate that it is a powerful tool to study
the structure and evolution of the galaxy together with the theoretical
evolutionary population synthesis methods (for an application of a
similar technique with, but with fewer colors, to moderate redshifts,
see Abraham et al. 1999). For this purpose, we pick M81 as the first
application of this multi-color approach. 

M81 is an excellent candidate
because it is a nearby early-type Sab spiral galaxy at a distance of
3.6 Mpc and with an angular size of $\sim 26^{\prime}$. The angular
extent is large enough such that the disk and bulge regions are well
separated from the ground. 
It has been the subject of numerous previous studies providing
a wealth of information with which to compare the new metallicity and 
internal reddening distribution. The internal reddening has been studied
by Kaufman et al. (1987, 1989), Devereux et al. (1995), Ho et al. (1996) 
and Allen et al. (1997). The metallicity has been studied by Stauffer \& 
Bothun (1984), Garnett \& Shields (1987), Brodie \& Huchra (1991) and 
Perelmuter et al. (1995).
In this paper,
we present a further detailed study of M81 using the unique dataset
obtained from the BATC \footnote[1]{The Beijing-Arizona-Taiwan-Connecticut
Multicolor Sky Survey} multi-color sky survey.

The outline of the paper is as follows.  Details of observations
and data reduction are given in section 2. In section 3, we provide
a brief description of the  model, and analyze the evolution of
the integrated colors, color indices with age and metallicity. The
observed two-dimensional spectral energy distributions (SEDs) of M81
were analyzed using stellar population synthesis models of Bruzual \&
Charlot (1996). The distributions of metallicity, age and interstellar
reddening are given in section 4. In section 5, we discuss how different
star formation histories and stellar population synthesis models change
our results, and compare our results with previous studies.
. The conclusions are summarized in section 6.

\section{OBSERVATIONS AND DATA REDUCTION}

\subsection{CCD Image Observation}

The large field multi-color observations of the spiral galaxy M81 were
obtained in the BATC photometric system. The telescope used is the
60/90 cm f/3 Schmidt Telescope of Beijing Astronomical Observatory (BAO),
located at the Xinglong station. A Ford Aerospace 2048$\times$2048 CCD
camera with 15$\mu$m pixel size is mounted at the Schmidt focus of the
telescope. The field of view of the CCD is $58^{\prime}$ $\times $ $
58^{\prime}$ with a pixel scale of $1^{\prime\prime}.7$.  

The multi-color BATC filter system includes 15 intermediate-band filters,
covering the total optical wavelength range from 3000 to 10000{\AA}
(see Fan et al, 1996). The filters were specifically designed to avoid
contamination from the brightest and most variable night sky emission
lines. A full description of the BAO Schmidt telescope, CCD, data-taking
system, and definition of the BATC filter systems are detailed elsewhere
(\cite{Fan96}, \cite{Chen00}).   To study the age, metallicity and
interstellar reddening of M81, the images of M81 covering most part
of the optical body of M81 were accumulated in 13 intermediate band
filters with a total exposure time of about 51 hours from February
5, 1995 to February 19, 1997.  The CCD images are centered at ${\rm
RA=09^h55^m35^s.25}$ and DEC=69$^\circ21^{\prime}50^{\prime\prime}.9$
(J2000). The dome flat-field images were taken by using a diffuse plate in
front of correcting plate of the Schmidt telescope. For flux calibration,
the Oke-Gunn primary flux standard stars HD19445, HD84937, BD+262606
and BD+174708 were observed during photometric nights. The parameters of
the filters and the statistics of the observations are given in Table 1.

\begin{table}[ht]
\caption[]{Parameters of the BATC filters and statistics of observations}

\vspace {0.5cm}
\begin{tabular}{ccccccrr}
\hline
\hline
 No. & Name& cw(\AA)\tablenotemark{a}& Exp.(h)&  N.img\tablenotemark{b}
 & RMS.\tablenotemark{c} & Err.\tablenotemark{d} &
 SM.err. \tablenotemark{e}\\
\hline
1  & BATC02& 3894   & 04:52& 26 &0.021&  7.2& 0.5\\
2  & BATC04& 4546   & 01:00& 10 &0.017& 11.5& 0.7\\
3  & BATC05& 4872   & 03:05& 11 &0.016&  3.5& 0.2\\
4  & BATC06& 5250   & 03:03& 13 &0.016&  5.2& 0.3\\
5  & BATC07& 5785   & 03:12& 16 &0.011&  7.7& 0.5\\
6  & BATC08& 6075   & 02:14& 10 &0.016& 18.3& 1.2\\
7  & BATC09& 6710   & 02:46& 14 &0.018&  5.7& 0.4\\
8  & BATC10& 7010   & 02:19& 11 &0.019& 25.1& 1.6\\
9  & BATC11& 7530   & 03:00& 13 &0.017&  8.2& 0.5\\
10 & BATC12& 8000   & 04:50& 16 &0.014& 16.1& 1.0\\
11 & BATC13& 8510   & 04:45& 15 &0.019& 20.9& 1.4\\
12 & BATC14& 9170   & 05:10& 17 &0.021& 25.7& 1.7\\
13 & BATC15& 9720   & 11:00& 35 &0.019& 25.8& 1.7\\
\hline
\end{tabular}\\
\tablenotetext{a}{Central wavelength for each BATC filter}
\tablenotetext{b}{Image numbers for each BATC filter}
\tablenotetext{c}{Zero point error, in magnitude, for each filter
as obtained from the standard stars}
\tablenotetext{d}{Background errors before image smoothing}
\tablenotetext{e}{Background errors after image smoothing}
\end{table}

\subsection{ Image data reduction }

The data were reduced with standard procedures, including bias subtraction
and flat-fielding of the CCD images, with an automatic data reduction
software named PIPELINE 1 developed for the BATC multi-color sky survey
(Fan et al. 1996). The flat-fielded images of each color were combined
by integer pixel shifting. The cosmic rays and bad pixels were corrected
by comparison of multiple images during combination. The images were
re-centered and position calibrated using the HST Guide Star Catalogue.
The sky background of the images was obtained by fitting image areas
free of stars and galaxies using the method described in Zheng et
al. (1999). The absolute flux of intermediate-band filter images was
calibrated using observations of standard stars. Fluxes as observed
through the BATC filters for the Oke-Gunn stars were derived by convolving
the SEDs of these stars with the measured BATC filter transmission
functions (\cite{Fan96}). {\it Column} 6 in Table 1 gives the zero point
error, in magnitude, for the standard stars in each filter. The formal
errors we obtain for these stars in the 13 BATC filters is $\la 0.02$
mag. This indicates that we can define the standard BATC system to an
accuracy of  $\la 0.02$ mag.

After background subtraction, the standard deviation of the background
for each image is 2.0ADU. Because the signal-to-noise ratio decreases
from the center to the edge of the galaxy, we smoothed the images with a
boxcar filter. The window sizes of the boxcar were selected depending on
the ADU values of the BATC10 band image (7010{\AA}).  If the ADU value
was less than 100, the pixel was set to zero; if the value was higher
than 100, the pixel was adaptive-smoothed by boxcar filter of $N \times
N $ (cell size), where N={\rm min}($151/\sqrt {{\rm ADU}_{\rm BATC10}},
15$).  By this method, the images were smoothed depending on the S/N of
each cell. In the central area of M81, the original pixels were used,
whereas near the edge of M81 the mean value of multiple pixels (cells)
were used, as a result, the spatial resolution decreased from center to
outer edge. The background errors before and after smoothing are given
in the last two columns in Table 1.

Finally, the flux derived at each point of M81 is listed in Table 2
\footnote[2]{The full Table 2 and color versions of Figs. 1, 4a, 5 and 6
are available electronically.}
. The results provide a two-dimensional spectral energy
distributions (SED) for M81. The ADU number of each image were converted
into units of $10^{-30} {\rm erg\,s^{-1}\,cm^{-2}\,Hz^{-1}}$. As an
example, we present some SEDs for the different areas of M81 in this
paper. The table gives the following information: {\it Column 1} and
{\it Column 2} give the $(X, Y)$ positions of the photometric center
of the regions, in units of arcseconds. The coordinate system is
centered on the nucleus of the galaxy (${\rm RA=09^h55^{m}35^s25}$;
${\rm DEC=69^{\circ}21^{\prime}50^{\prime\prime}.9}$, in J2000).
The $X$-axis is along the E-W direction with positive values towards the
east, and the $Y$-axis is along the N-S direction with positive values
towards the north.  {\it Column} 3 to {\it Column} 14 give the fluxes
relative to the BATC08 filter (6075{\AA}). {\it Column} 15 gives the
flux in the  BATC08 filter in units of ${\rm 10^{-30} erg s^{-1} cm^{-2}
Hz^{-1}}$. For convenience in later discussions, we define the `central'
areas as regions with $\DeltaRA \leq 1^{\prime}.0, \DeltaDec \leq
1^{\prime}.9$, the `bulge' region with $1.0< \DeltaRA \leq 3^{\prime}.4,
1.9 <\DeltaDec \leq 4^{\prime}.0$ and the `disk' region with $3^{\prime}.4
<\DeltaRA \le 6^{\prime}.2, 4^{\prime}.0 <\DeltaDec <8^{\prime}.0$. 

\begin{deluxetable}{llrrrrrrrrrrrrc}
\tablewidth{0pc}
\tablecaption{Two dimensional spectral energy distributions
for the central, bulge and disk areas of M81}
\tablehead{ \colhead{}&\colhead{}&
\multicolumn{12}{c}{Fluxes in different filters relative
to the $F_{BATC08}$ filter}& \colhead{Flux of}\\
\cline{3-14}
\colhead{X}&\colhead{Y}&\colhead{02}&\colhead{04}&
\colhead{05}&\colhead{06}&\colhead{07}&
\colhead{09}&\colhead{10}&\colhead{11}&\colhead{12}&
\colhead{13}&\colhead{14}&\colhead{15}&\colhead{BATC08} }
\startdata

292&330&0.083&0.287&0.368&0.441&0.660&0.726&0.864&1.104&1.227&1.303&1.604&1.717&9232\nl

291&330&0.084&0.289&0.367&0.440&0.661&0.723&0.862&1.096&1.216&1.287&1.594&1.703&8923\nl

290&330&0.085&0.294&0.366&0.444&0.665&0.722&0.865&1.094&1.221&1.278&1.585&1.702&8527\nl

292&331&0.085&0.289&0.366&0.443&0.665&0.720&0.866&1.102&1.221&1.301&1.604&1.727&8695\nl

291&331&0.086&0.291&0.365&0.444&0.669&0.726&0.872&1.101&1.224&1.304&1.604&1.726&8398\nl

\nl
286&399&0.098&0.309&0.361&0.451&0.660&0.729&0.866&1.073&1.213&1.244&1.571&1.683&1199.\nl

287&400&0.099&0.307&0.360&0.450&0.656&0.721&0.864&1.068&1.211&1.233&1.568&1.665&1168.\nl

284&401&0.099&0.313&0.365&0.454&0.659&0.728&0.863&1.065&1.198&1.233&1.556&1.654&1179.\nl

285&402&0.102&0.315&0.365&0.454&0.660&0.724&0.862&1.063&1.194&1.227&1.547&1.646&1154.\nl

283&403&0.105&0.320&0.374&0.462&0.671&0.736&0.872&1.068&1.202&1.228&1.546&1.644&1150.\nl

\nl
302&499&0.207&0.415&0.442&0.526&0.712&0.743&0.886&1.009&1.161&1.186&1.445&1.498&266.\nl

300&499&0.217&0.427&0.457&0.534&0.723&0.751&0.894&1.014&1.150&1.181&1.454&1.489&277.\nl

299&500&0.218&0.429&0.457&0.533&0.719&0.743&0.891&1.009&1.139&1.173&1.441&1.482&278.\nl

299&502&0.219&0.431&0.459&0.538&0.723&0.732&0.899&1.018&1.142&1.174&1.438&1.496&268.\nl

298&502&0.221&0.436&0.463&0.542&0.725&0.736&0.904&1.024&1.146&1.177&1.453&1.511&271.\nl

\enddata
\tablecomments{Rows 1-5, 6-10, 11-15 are for the central,
bulge and disk areas respectively. The absolute flux (last column)
is in units of $10^{-30} {\rm erg s^{-1} cm^{-2}  Hz^{-1}}$. The full
table for M81 is available electronically. The  pixel values for the
center
of M81 is (300, 375). }
\end{deluxetable}

We show a black and white image of M81 in Figure 1. To display the
features clearer, a ``true-color'' image of M81 is available in the
electronic version, which is combined with the
``blue''(3894{\AA}), ``green''(5785{\AA}) and ``red''(7010{\AA}) filters.
The filters selected here are free from any strong emission lines. From
this image, we can see directly the stellar population difference in
different areas of the spiral galaxy M81.  In the following sections,
we will analyze quantitatively the stellar populations in M81 with our
13 color data.

The bright HII regions consist of young clusters, and the 
evolutionary population synthesis methods used in this paper
do not represent young clusters well. In addition, the central 
nucleus of M81 exhibits some of the same characteristics as classical 
Seyfert galaxies, has no evidence of stellar clusters or a population 
of hot young stars (\cite{kaufman96}, \cite{devereux97}, \cite{davidge99}).
So we mask some bright HII regions and the central nucleus of M81 
(shown as white spots in Fig. 4a, 5, 6, including the foreground stars) 
in this paper. 

\begin{figure}
\centerline{\psfig{file=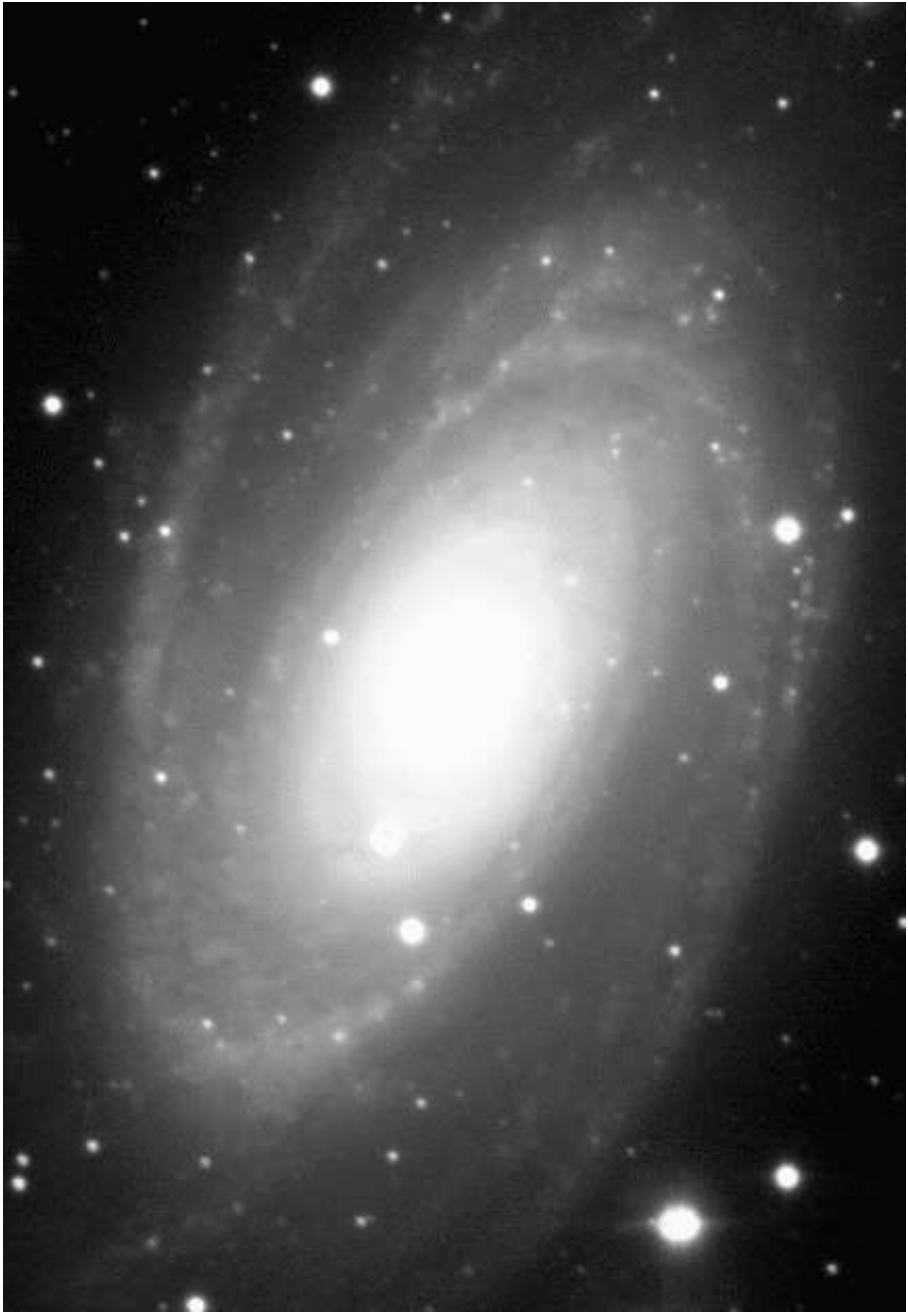,width=12.0cm,angle=0}}
\caption{``True-color" estimate of the M81 generated by using
the BATC02 (3894{\AA}) filter image for blue, BATC07 (5785{\AA}) for
green, and BATC10 (7010{\AA}) for red; the image is balanced by making
the background old population orange and hot stars blue.  The center
(origin) of the image is located at $\rm \alpha=09^h55^{m}35^s.25, \delta
=69^{\circ}21'50''.9 (J2000.0)$. The image size is about $17^{\prime}.0$
by $21^{\prime}.0$.  North is up and east is to the left. We refer to
the region with $\DeltaRA \leq 1^{\prime}.0, \DeltaDec \leq 1^{\prime}.9$
as the central region, the region with $1.0< \DeltaRA \leq 3^{\prime}.4,
1.9 <\DeltaDec \leq 4^{\prime}.0$ as the bulge, and the region with
$3^{\prime}.4 <\DeltaRA \le 6^{\prime}.2, 4^{\prime}.0 <\DeltaDec
<8^{\prime}.0$ as the disk. A black and white image of Figure 1 be
displayed in paper, the ``true-color'' image of Figure 1 is available
in the electronic version.}
\label{fig1}
\end{figure}

\section{DATABASES OF SIMPLE STELLAR POPULATIONS}

A simple stellar populations (SSP) is defined as a single generation
of coeval stars with fixed parameters such as metallicity, initial
mass function, etc (\cite{Buzzoni97}). In evolution synthesis models,
they are modeled by a collection of stellar evolutionary tracks with
different masses and initial chemical compositions, supplemented
with a library of stellar spectra for stars at different evolutionary
stages. Because SSPs are the basic building blocks of synthetic spectra
of galaxies that can be used to infer the formation and subsequent
evolution of the parent galaxies (\cite{Jab96}). In order to study the
integrated properties of stellar population in M81, as the first step,
we use the SSPs of Galaxy Isochrone Synthesis Spectra Evolution Library
(Bruzual \& Charlot 1996 hereafter GSSP). We study the SSPs as the
first step for two reasons. First, they are simple and reasonably well
understood, so it is important to see what one can learn using this
simplest assumption, and then check whether more complex star formation
history give qualitatively similar conclusions. This is a common approach
often taken in the evolutionary population synthesis models for galaxies
(\cite{Vazdekis97}, \cite{mayya95}). Second, although we assume each pixel 
is described by an SSP, we emphasize that the whole galaxy is not SSP; so 
our assumption is not as strong as it may seem. Nevertheless, this is a 
significant assumption. Fortunately, it appears the adoption of more 
complex star formation history does not change the results 
qualititatively; we return to this issue in \S 5.1.
 
\subsection{Spectral Energy Distribution of GSSPs}

The Bruzual \& Charlot (1996) study has extended the Bruzual \& Charlot
(1993) evolutionary population synthesis models. The updated version
provides the evolution of the spectrophotometric properties for a wide
range of stellar metallicity. They are based on the stellar evolution
tracks computed by Bressan et al. (1993), Fagotto et al.  (1994), and
by Girardi et al. (1996), who use the radiative opacities of Iglesias
et al. (1992). This library includes tracks for stars with metallicities
$Z=0.0004, 0.004, 0.008, 0.02, 0.05,$ and $0.1$, with the helium abundance
given by $Y=2.5Z+0.23$ (The solar metallicity is $Z_\odot=0.02$). The
stellar spectra library are from Lejeune et al. (1997,1998) for all
the metallicities listed above, which in turn consist of Kurucz (1995)
spectra for the hotter stars (O-K), Bessell et al. (1991) and Fluks
et al. (1994) spectra for M giants, and Allard \& Hauschildt (1995)
spectra for M dwarfs. The initial mass function is assumed to follow the
Salpeter's (1955) form, $dN/dM \propto M^{-2.35}$, with a lower cutoff
$M_{\rm l}=0.1M_{\odot}$ and an upper cutoff $M_{\rm u}=125M_{\odot}$
(\cite{Sawicki98}).

\subsection{Integrated Colors of GSSPs}

To determine the age, metallicity and interstellar medium reddening
distribution for M81, we find the best match between the observed colors
and the predictions of GSSP for each cell of M81. Since the observational
data are integrated luminosity, to make comparisons, we first convolve
the SED of GSSP with BATC filter profiles to obtain the optical
and near-infrared integrated luminosity.  The integrated luminosity
$L_{\lambda_i}(t,Z)$ of the $i$th BATC filter can be calculated with
\beq 
L_{\lambda_i}(t,Z) =\frac{{\int_{\lambda_{\rm min}(i)}^{\lambda_{\rm
max}(i)} F_{\lambda}(t,Z)\varphi_i(\lambda)d\lambda}} {{\int_{\lambda_{\rm
min}(i)}^{\lambda_{\rm max}(i)} \varphi_i(\lambda)d\lambda}}, 
\eeq
where the $F_{\lambda}(t,Z)$ is the spectral energy distribution of
the GSSP of metallicity $Z$ at age $t$, $\varphi_i(\lambda)$ is the
response functions of the BATC filter system, and $\lambda_{\rm min}(i)$
and $\lambda_{\rm max}(i)$ are respectively the maximum and the minimum
effective wavelength of the $i$th filter ($i=1, 2, \cdot\cdot\cdot, 13$).

The absolute luminosity can be obtained if we know the distance to a
galaxy and the extinction along the line of sight. Since we do not
know the exact distance to M81, in this paper, we shall work with
the colors that are independent of the distance. We calculate
the integrated colors of a GSSP relative to the BATC filter BATC08
($\lambda=6075${\AA}):

\beq 
\label{color}
C_{\lambda_i}(t,Z)={L_{\lambda_i}(t,Z)}/{L_{6075}(t,Z)}.  
\eeq

As a result, we obtain intermediate-band colors for 6 metallicities from
$Z=0.0004$ to $Z=0.1$. In the panels of Fig. 2, we plot the colors as
a function of age for GSSP with different metallicities. The following
remarks can be made. (a) It is apparent that there is a uniform tendency
for SSPs to become redder for all colors as the metallicity increases
from $Z=0.0004$ to $Z=0.05$.  The near-UV and optical colors show the same
qualitative behavior as those at longer wavelengths. (b) There is a wide
range in age (from 1 to 20 Gyr) in which the colors vary monotonically
with time except for the highest metallicity $Z=0.1$. Therefore, once we
know the metallicity and interstellar reddening, we can use these colors
to determine the age distribution of M81, provided that the stellar
population is well modeled by SSPs. (c) For $Z=0.1$, there is only a
limited age range for the monotonic behavior in colors. One reason for
this behavior is the appearance of AGB-manqu\'{e} stars at $Z=0.1$.
These stars skip the AGB phase and directly go through a long-lived
hot HB phase (\cite{Bruzual96}).  There are very few, if any, examples
of galactic stars with such a high metallicity. So our results are not
affected by this peculiar high-metallicity case.

\begin{figure}
\figurenum{2}
\centerline{\psfig{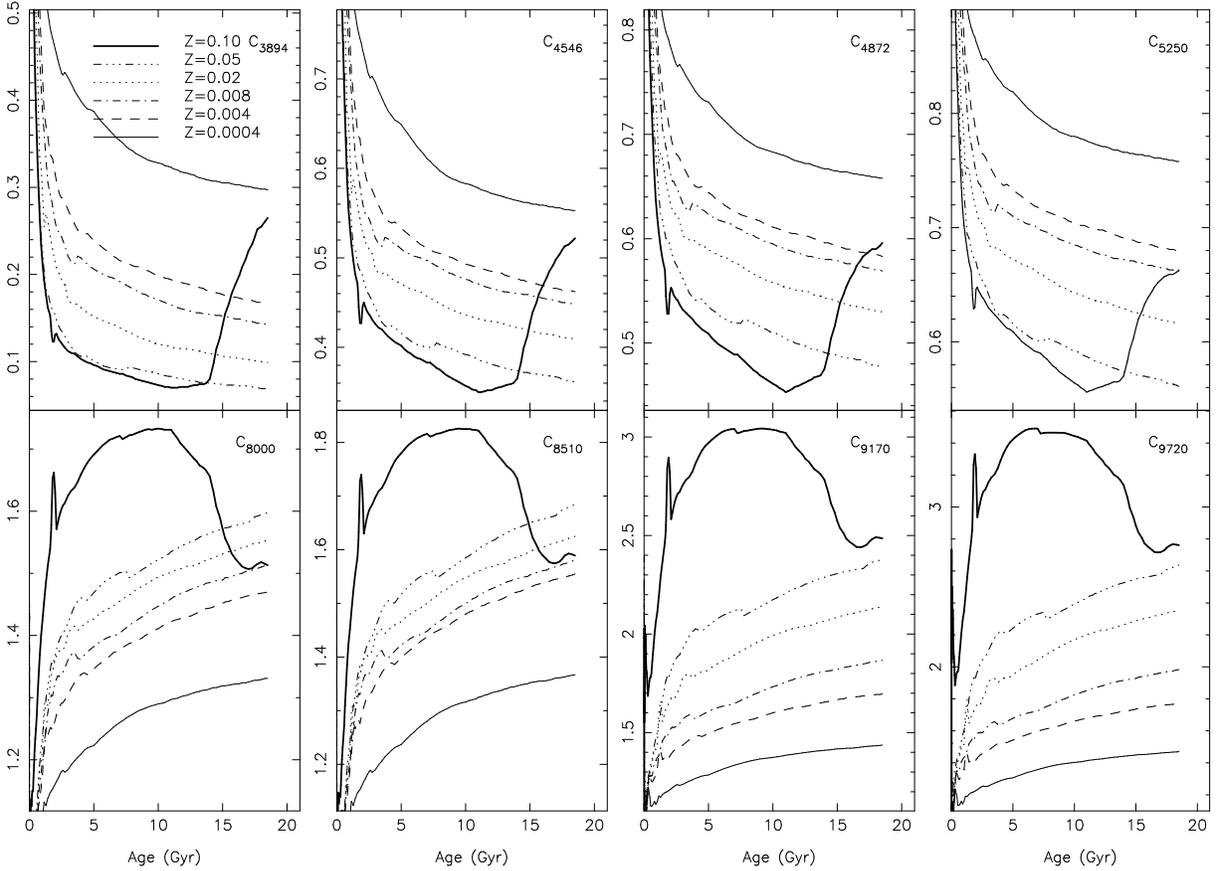}}
\caption[fig2.ps]{Evolution of eight selected colors $C_{\lambda_i}$
(cf. eq. \ref{color}) for simple stellar population (SSP) models for
different metallicities as predicted by the GISSEL96 library. Each
line pattern represents a different metallicity, as indicated inside
the frame. All the models shown are computed with a the Salpeter(1955)
initial mass function  with a lower cutoff of $0.1M_\odot$ and an upper
cutoff of $125M_\odot$.}
\label{fig2}
\end{figure}

\subsection{Color Indices of GSSPs}

The observed colors are affected by interstellar reddening, which will of
course complicate our interpretations (\cite{Ostlin98}).  The interstellar
reddening in the center region of M81 can be measured by its emission
lines, but for the outer regions, the problem becomes very complex. If
we suppose that the extinction law from 3800{\AA} to 10000{\AA} has
no high frequency features, the spectral indices will not be affected
much by the uncertainties in the extinction, so we can use the spectral
indices to reduce the effect of interstellar extinction.

Spectral indices are (by definition) constructed by means of a
central band pass and two pseudo-continuum band-pass on either side
of the central band. The continuum flux is interpolated between the
middle-points of the pseudo-continuum band passes (\cite{Bressan96},
\cite{Worthey97}). Since we only observed M81 with intermediate-band
filters, not a genuine one-dimensional spectrum, so we must use some
pseudo-color indices to replace the conventional definitions.  We define
a color index $I_{\lambda_j}(t,Z)$ of a SSP by
\beq 
I_{\lambda_j}(t,Z)=L_{\lambda_j}(t,Z)/L_{\lambda_{j+1}}(t,Z), 
\eeq
where $L_{\lambda_j}(t,Z)$ is the luminosity of a SSP with the metallicity
$Z$ at age $t$ and wavelength $\lambda_j$, $L_{\lambda_{j+1}}(t,Z)$ is
the luminosity in the $(j+1)$th filter for the same SSP. The color indices
can reduce the effect of reddening, especially in the wavelength region
longer than 5000{\AA}.

Among all the BATC filter bands, we find that the color index centered
at 8510{\AA} ($I_{8510}$) is much more sensitive to the metallicity than
to the age; the center of this filter band is near the CaII triplets
($\lambda=8498, 8542, 8662{\rm \AA}$).  The strength of CaII triplet
depends on the effective temperature, surface gravity and the metallicity
for late type stars (\cite{Zhou91}).  

In an old stellar system, the effect of metallicity on CaII triplet
becomes prominent. In fact, we find that there is a very good relation
between the flux ratio of $I_{8510}\equiv L_{8510}/L_{9170}$ and the
metallicity for stellar populations older than 1 Gyr.  We plot this
relation in Figure 3. Similar  relations are also found in many other
observation and stellar population synthesis models.

The relation shown in Figure 3 is crucial for our metallicity
determination and later studies, so it is important to check whether this
is indeed a reliable method.  An early study by Alloin \& Bica (1989),
based on the analysis of stars, star clusters and galaxy nuclei indicated
a strong correlation of CaII triplet with the surface gravity, $\log
g$. However, further studies suggested that the CaII triplet strength
depends not only on the surface gravity but also on the metallicity.
A detailed analysis of the behavior of the CaII triplet feature as a
function of stellar parameters was performed by Erdelyi-Mendes \& Barbuy
(1991), making use of a large grid of synthetic spectra. They concluded
that CaII triplet has a weak dependence on the effective temperature,
a modest dependence on surface gravity, but a quite important dependence
on metallicity.  They even suggested that the CaII triplet strength
may vary exponentially with the metallicity.  Moreover, these lines
have been studied by Diaz et al. (1989) , Mallik (1994) and Idiart et
al. (1997). They have also suggested the CaII triplet strengths depend
on the metallicity (Mayya 1997). So the relation between the $I_{8510}$
and metallicity seems to be reliable and can be used to determine the
metallicities; our own investigation of GSSPs seems to be consistent with
these recent studies.

\begin{figure}
\figurenum{3}
\centerline{\psfig{file=kongxu.fig3a.ps,width=12.0cm,angle=-90}}
\end{figure}

\begin{figure}
\figurenum{3}
\centerline{\psfig{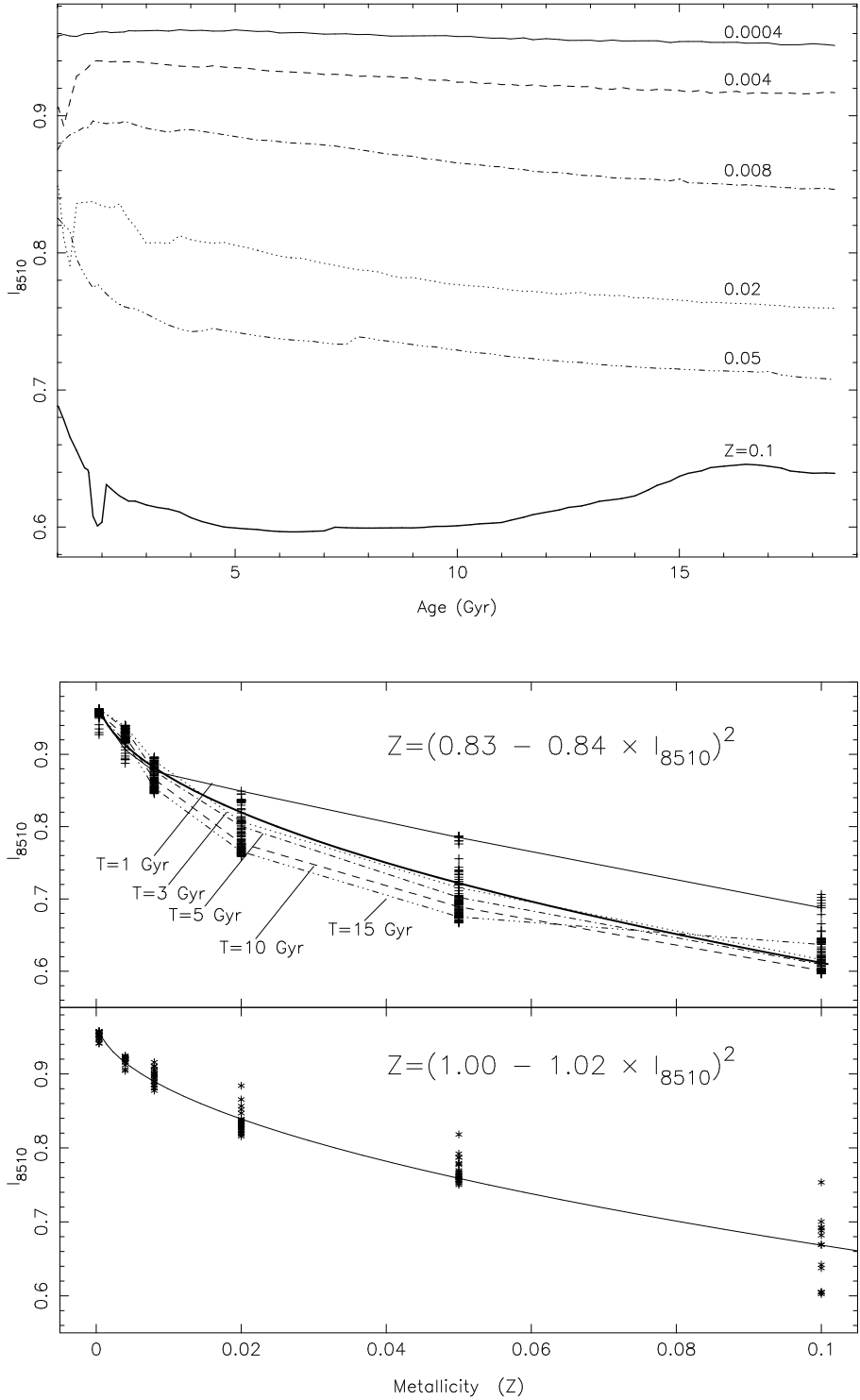}}
\caption[kongxu.fig3.ps]{a) Dependence of the pseudo color-indice $I_{8510}$
with metallicity for the GISSEL96 SSP models.  The line symbols are the
same as in Fig. 2. The metallicities are labeled above each line.  b) The
correlation between the pseudo color-index $I_{8510}$ and metallicity,
for five different ages ($T=15,10,5,3,1$ Gyr). The solid line shows the
best fit through the curves. The lower panel for the PSSPs, the upper
panel for the GSSPs.
\label{fig3}}
\end{figure}

\section{DISTRIBUTION OF METALLICITY, AGE AND REDDENING}

In general, the SED of a stellar system depends on age, metallicity
and reddening along the line of sight. The effects of age, metallicity
and reddening are difficult to separate (e.g., \cite{Calzetti97},
\cite{Origlia99},\cite{Vazdekis97}). Older age, higher metallicity or
larger reddening all lead to a redder SEDs of stellar systems in the
optical (\cite{Molla97}, \cite{Bressan96}). In order to separate the
effects of age, metallicity and interstellar reddening of M81, we first
determine the metallicity by the color index $I_{8510}$, as discussed
above, and then obtain the age and reddening by using GSSP model of
known metallicity and a extinction law (see \S 4.2).  

\subsection{Metallicity distribution}

As discussed in \S 3.3, there is a good correlation between the color
index $I_{8510}$ and the metallicity. We will use the relation obtained
from GSSP; this relation is similar for other stellar population synthesis
models. We find that the correlation can be fit with a simple formula: 
\beq 
Z=(0.83 - 0.84 \times I_{8510})^{2}.
\eeq
This curve is shown in the upper panel of Figure 3. The scatter in Figure
3 is due to the difference in age. If the ages are younger than 1 Gyr,
the scatter becomes larger. Figure 3 shows that for a stellar system of
ages older than 0.5 Gyr, we can estimate the metallicity  with an error
less than the interval of metallicity given by GSSPs. 

Using this method we obtained the metallicities for each part of 
M81 except for the nucleus and the H$\alpha$ line emission region 
(we have masked them out, see \S 2.2).  Figure 4a shows the resulting 
metallicity map of M81. 
Figure 4b shows the radial distribution of the metallicity, the
curve is derived from the Fig. 4a by averaging over ellipses
of widths $17\arcsec$ along the major axis. We used an
inclination angle of $i=59^{\circ}$ and a position angle of
${\rm PA}=157^{\circ}$ for the major axis of the galaxy.

To our surprise, we do not find, within our errors, any obvious
metallicity gradient from the central region to the bulge and disk
of M81.  In most parts of M81, the mean metallicity is about 0.03 with
variation $\lesssim 0.005$. 
These results are identical to past suggestions that early-type spirals
may have relative high abundances and weak gradients.
Taking into account of the age scatter,
the true value of the metallicity is likely within a range between
$Z =0.02$ and $Z =0.05$. From the metallicity map of M81, we can also
clearly see that in some outer regions the metallicities are higher;
most of these regions are located in spiral arms and around HII regions,
where a younger stellar population is present.

\begin{figure}
\figurenum{4a}
\centerline{\psfig{file=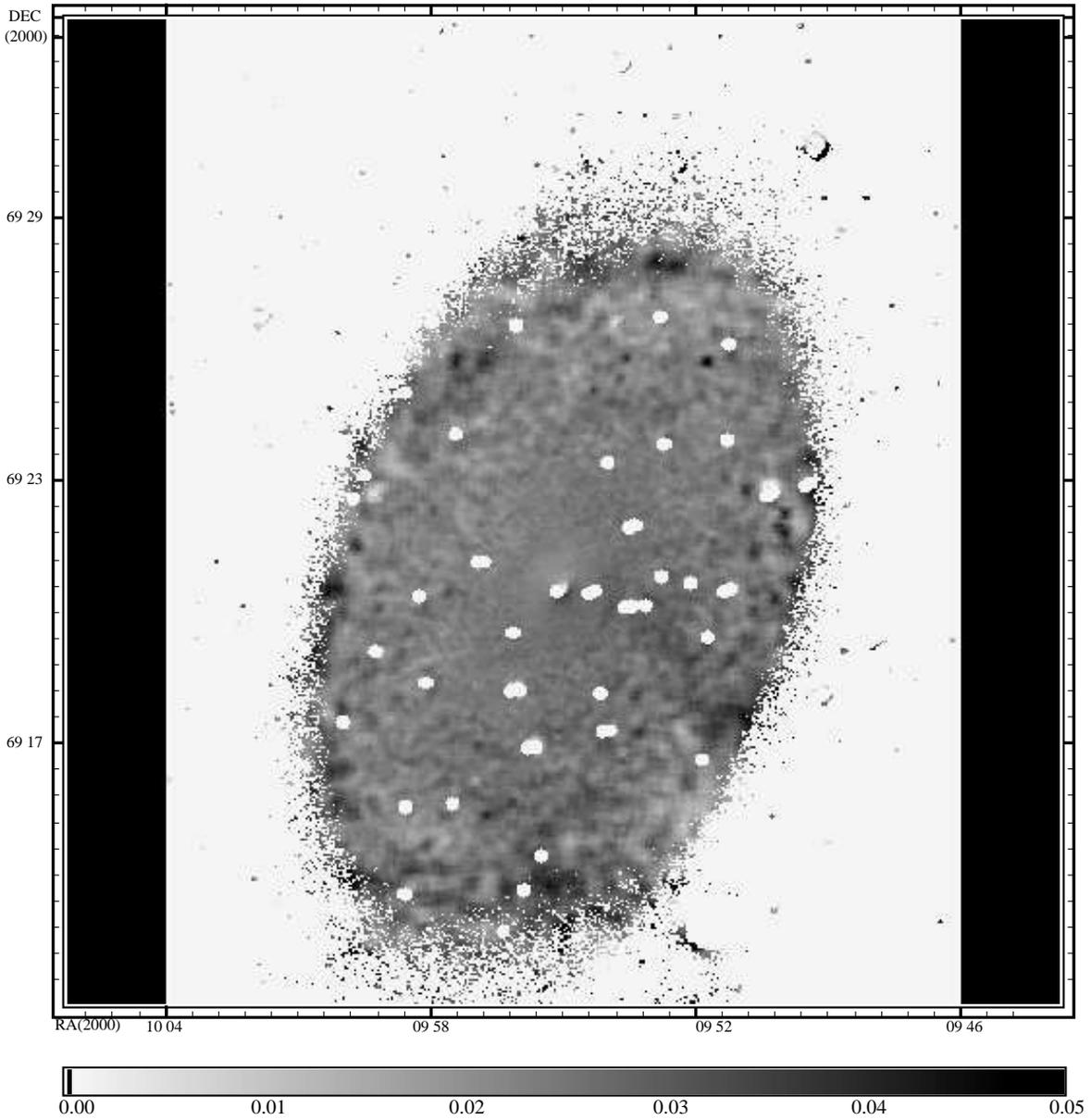,width=16.0cm}}
\caption[kongxu.fig4a.ps]{A map of the population metallicity in M81.
Light gray corresponds to the low and dark gray to the high  metallicity.
Some bright HII regions, the central nucleus of M81 and most of
foreground stars are masked, shown as white spots.
\label{fig4a}}
\end{figure}

\begin{figure}
\figurenum{4b}
\centerline{\psfig{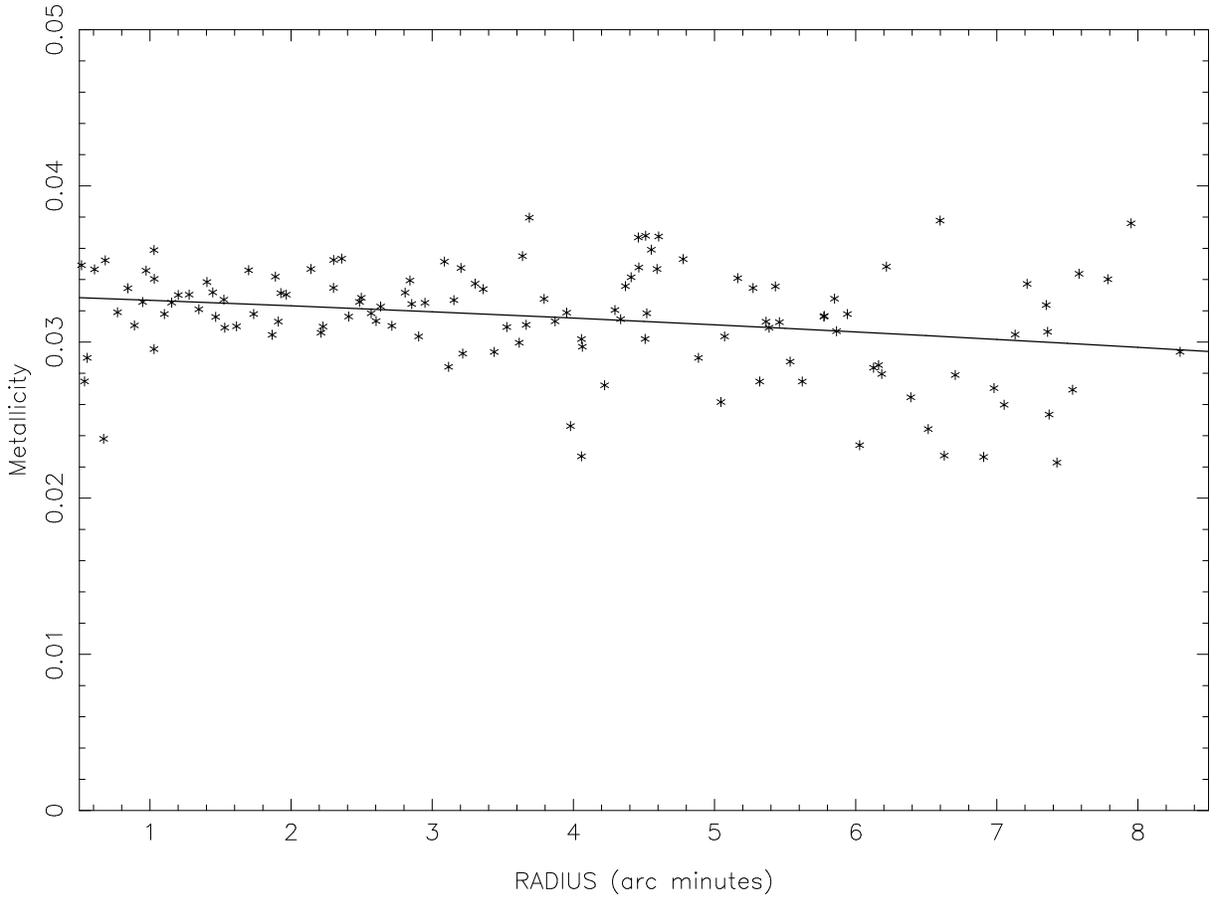}}
\caption[kongxu.fig4b.ps]{The radial distribution of metallicity
for M81, averaging over ellipses along the major axis (its position
angle is taken to be $157^{\circ}$). The line is the regression line
obtained by a polynomial fit.
No obvious metallicity gradient exists from the
central region to the bulge and disk of M81.
\label{fig4b}
}
\end{figure}

\subsection{Age and reddening distribution}

Since we model the stellar populations by SSPs, the observed colors for
each cell are determined by two parameters: age, $t$, and dust reddening,
$E(B-V)$.  In this section, we will determine these parameters for M81
simultaneously by a least square method. The procedure is as follows.
For given reddening and age  (recall that the metallicity is known, see
the previous subsection), we can obtain the predicted integrated colors
by convolving the dust-free predictions from GSSP with the extinction
curve given by Zombeck (1990). The best fit age and reddening values
are found by minimizing the difference between the observed colors and
the predicted values:
\beq 
R^2(x,y,t,Z,E)=\sum_{i=1}^{12}[C_{\lambda_i}^{\rm
obs}(x,y)-C_{\lambda_i}^ {\rm ssp}(t, Z, E)]^2, 
\eeq
where $C_{\lambda_i}^{\rm ssp}(t, Z, E)$ represents integrated color in
the $i$th filter of a SSP with age $t$, metallicity $Z$ and reddening
correction $E$, and $C_{\lambda_i}^{\rm obs}(x,y)$ is the observed
integrated color at position $(x,y)$.

Figure 5 shows the age map for M81. It clearly indicates that the
stellar population in the central regions is much older than that in
the outer regions and the youngest components reside in the spiral arms
of M81. There is a smooth age gradient from the center of the galaxy to
the edge of the bulge. The age in the innermost central region (within
17\arcsec) is older than $\ga$ 15 Gyr.  The age at the more extended central
region is about 9 Gyr. In the bulge edge area, the age is about 4 Gyr. In
contrast, the stellar component in the disk area is much younger than
that in the bulge region. The mean age in disk area is about 2.0 Gyr. We
can see that the age in spiral arms is even younger than the inter-arm areas,
about 1 Gyr.

Because the age obtained in the outer disk region is around 1 Gyr, the
metallicity, which is determined by the color indices, might have large
errors (see \S 4.1). The errors in turn will make the age determination
uncertain. Therefore the age for disk can only be regarded as a rough
estimation. However, the general trend in the age distribution should
be reliable.

\begin{figure}
\figurenum{5}
\centerline{\psfig{file=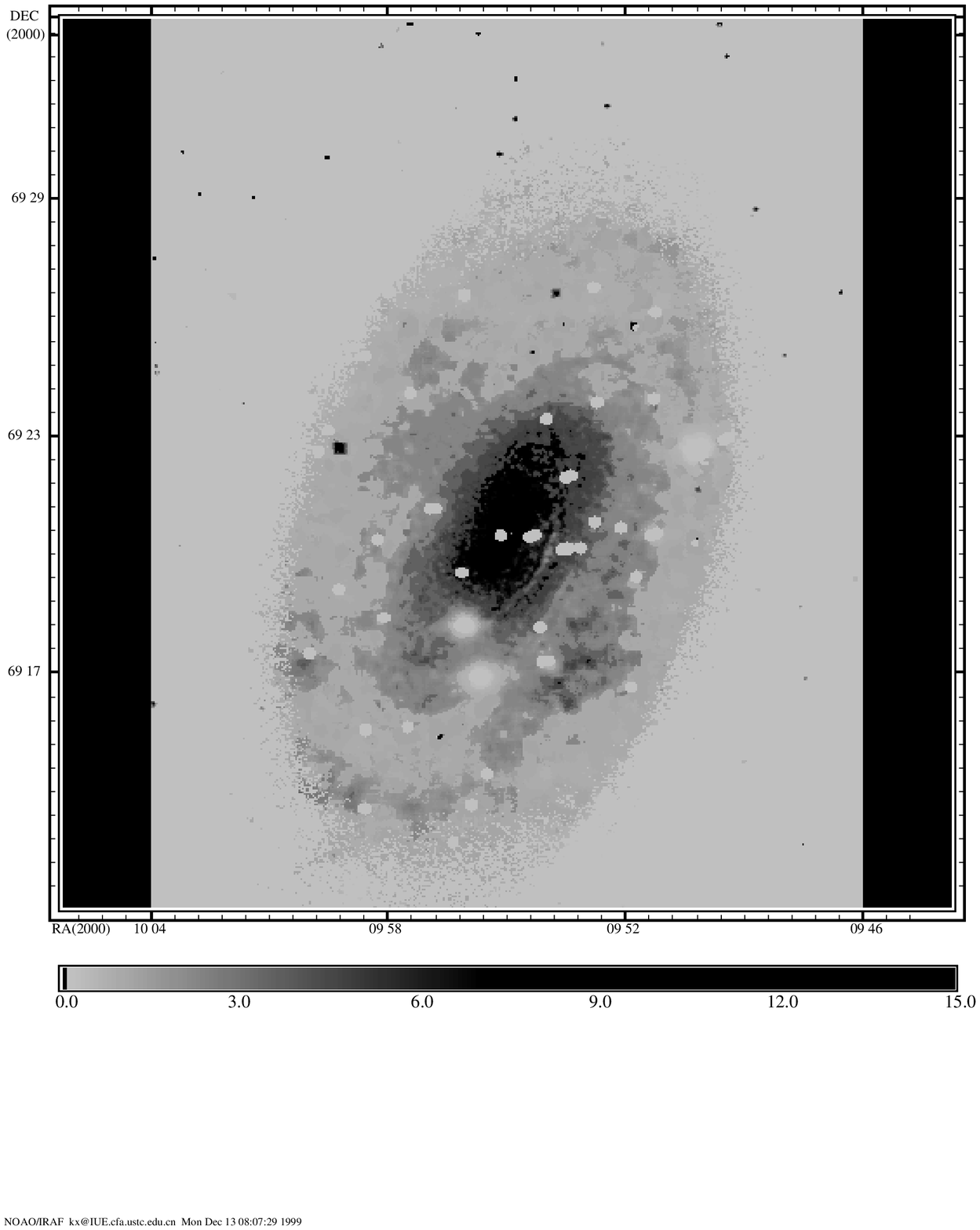,width=16.0cm}}
\caption[kongxu.fig5.ps]{A map of the population ages in M81.
Light gray corresponds to the young and dark gray to the oldest zones.
Same as in Fig. 4a, the white spots represent the masked regions,
such as bright HII regions, the central nucleus of M81 and most
of foreground stars.
\label{fig5}}
\end{figure}

Figure 6 shows the reddening map for M81. From this figure, we find
a large difference in reddening between the bulge and the disk. In the
bulge, the reddening, E(B-V), is in the range of 0.08 to 0.15. For regions
where E(B-V) is larger than 0.1, we found some spiral like cirrus. There
is a very obvious high reddening lane around the nucleus with reddening
equal to or higher than the disk area.  The maximum of E(B-V) in the lane
reaches 0.25. This half-loop lane can be verified in the future with IR
or CO line observations. In the disk area, the mean reddening of E(B-V)
is about 0.20. 
In the central regions, the mean reddening of E(B-V) is very small. These
results suggest that dust is largely absent in the central regions
of M81. The dust seems to be distributed mainly along the inner part
of spiral arms and around the nucleus.  Color versions of Figs. 4-6 are
available electronically.

\begin{figure}
\figurenum{6}
\centerline{\psfig{file=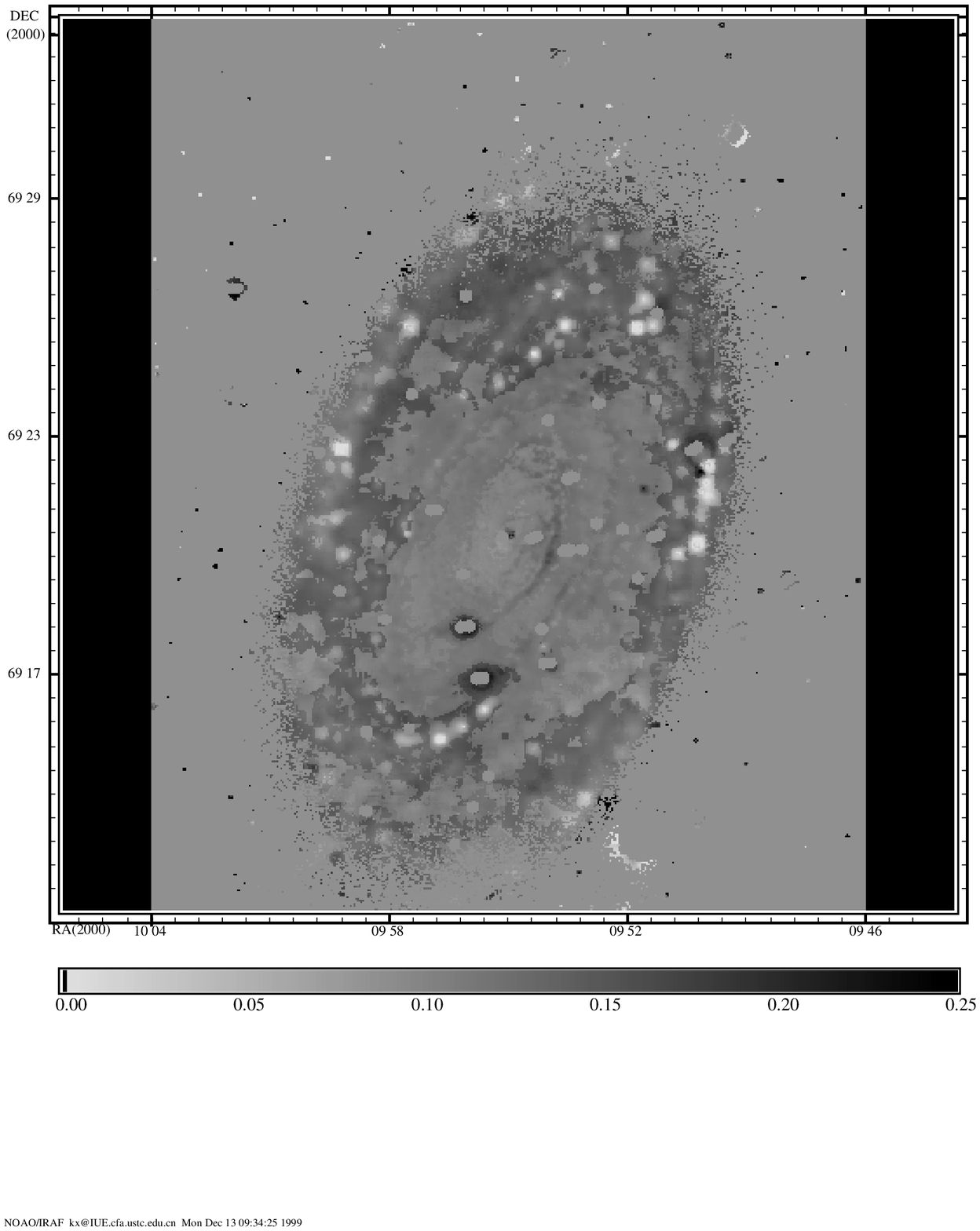,width=16.0cm}}
\caption[kongxu.fig6.ps]{The interstellar
reddening map of M81, using the method described in the text.
\label{fig6}
}
\end{figure}

\section{DISCUSSION}

The results we presented so far are based on  the (strong) assumption that
all stars in a small region form in an instantaneous burst and hence the
stellar population of each cell can be modeled as SSPs. Unfortunately, the
star formation rate history is an essential but very uncertain ingredient
in the evolutionary population synthesis method since it can vary from
galaxy to galaxy and from region to region inside a single galaxy as
well. It is only for simplicity that we have adopted an instantaneous star
formation history, clearly it is important to check whether the results
are significantly changed if one varies the star formation history; we
address this issue in \S 5.1.  While there seems to be general agreement
between the GISSEL96 SSPs models (which we used) with other similar ones
(\cite{Charlot96}), there are some fine differences. In \S 5.2, we study
how the results are changed if we adopt a different population synthesis
model from the Padova group. In \S 5.3, we compare our results with
earlier works.

\subsection{Continuous Star Formation}

We assume that stars are formed from the interstellar gas exponentially
with a characteristic time-scale $\tau$, i.e., $\Psi(t)=\Psi_0
\exp^{-t/\tau}$. This model is often used to calculate the integrated
colors of galaxies (\cite{Kennicutt98}) and allows a more diverse star
formation history. If $\tau\rightarrow\infty$ the model approximates
constant star formation rate, while for $\tau\rightarrow 0$ it
approximates an instantaneous burst (\cite{Abraham99}). It seems that
spiral galaxies could be well fitted with $\tau$ of the order of several
Gyr (Fioc et al. 1997).

Since we do not know the appropriate $\tau$ value for M81, we have
explored the values $\tau = 0.1, 1, 3$ Gyr.  We have calculated the
age, metallicity and interstellar reddening distributions for each
value of $\tau$. As the value of $\tau$ increases, the age ($t$)
increases throughout M81, while the interstellar reddening decreases
and the metallicity is little changed. Although the numerical values
of these quantities do change in each point of M81, the two dimensional
distributions of age, metallicity and interstellar reddening of M81 for
different values of $\tau$ are similar to the ones shown in Figs. 4-6.

\subsection{Comparison With Other SSP Models}

There exist a number of SSP models that are synthesized with different
approaches. It is important to check the sensitivity of our results
to the SSP models adopted.  The SSP models from the Padova group
(hereafter, PSSPs) is suitable for the comparison, because they use a
similar technique of ``isochrone synthesis'' to predict the spectral
evolution of stellar populations.  The PSSPs provide the basis for the
population synthesis models (\cite{Bressan94}, see Bressan et al. 1996
for revisions and extensions). The PSSPs use a comprehensive set of
stellar evolutionary tracks of the Padova group for a wide range of
initial chemical compositions from $Z = 0.0004$ to $Z = 0.1$ with ${\Delta
Y}/{\Delta Z} = 2.5$.  The initial masses of the evolutionary tracks cover
the range of $0.6 - 120 M_{\odot}$, except for the set of metallicity
Z=0.1, where the masses are from $0.6 - 9 M_{\odot}$.  The initial
mass function is the Salpeter (1955) law. More details can be found in
Bressan et al. (1994), Silva (1995), and Tantalo et al. (1996). The main
difference between GSSPs and PSSPs is the library of stellar spectra:
the GSSPs use theoretical stellar spectra from   Lejeune et al. (1997)
while the PSSPs use theoretical stellar spectra from Kurucz (1992).

Using our method, we calculate the colors and color indices for each
PSSP in the BATC filter system. Using similar procedures as for the GSSP
(see \S 4), we obtained the metallicity, age and reddening distributions
of M81. For PSSP, the metallicity map of M81 again has no obvious
metallicity gradient, but the mean metallicity is somewhat higher, about
0.035. There is a smooth age gradient from the center of M81 to the edge
of the bulge, except that the mean age in the disk area is lower than 1
Gyr. The interstellar reddening value from PSSP is obvious bigger than
that from GSSP. In the bulge, the reddening value is in  the range of
0.18 to 0.35. In the disk area, the mean reddening of E(B-V) is about
0.40. The reddening value in the central region amount to 0.15. The
distributions of metallicity, age and interstellar reddening are very
similar to those found using GSSP.

\subsection{Comparison with previous work}

Numerous determinations of the amount of extinction in M81 have recently
been obtained.
Allen et al. (1997) compared the detailed distribution of HI, $H\alpha$,
150 nm Far-UV continuum emission in the spiral arms of M81. They found every
reliable bright peak in the $H\alpha$ has a peak in the Far-UV, and concluded
that the effects of extinction on the morphology are small 
on the spiral arms. Filippenko \& Sargent (1988), based on the ratio of the 
narrow components of $H\alpha$ and $H\beta$, concluded that the central 
regions of M81 is reddened by E(B-V)=0.094 mag. These results are very 
similar to our results for internal reddening, that the mean reddening in
the spiral arms and in the central regions of M81 are small. 
In addition, Kaufman et al. (1987, 1989), using the $H\alpha$ and radio 
continuum (\cite{bash86}) 
observations, have studied the distribution of extinction 
along the spiral arms in M81, and obtained a mean $A_v = 1.1 \pm 0.4$ 
mag for 42 giant HII regions with high surface brightness.
Hill et al. (1992) used Near-UV, Far-UV and V-band images of M81 and cluster 
models to derive an $A_v =1.5$ mag  for the HII regions on the arms.
These internal reddening values are larger than ours for
spiral arms. It can be explained for two reasons. First, the internal
reddening value of Kaufman et al. are derived from the giant HII 
regions of M81 with high surface brightness, but these bright HII 
regions are excluded in our study. So we have have used different regions
for the internal reddening study, it is therefore not clear that they
must agree.
Second, the continuum regions are affected less by reddening than the 
emission line regions in a galaxy. The internal reddening value of 
Kaufman et al. is the reddening for emission line regions, but 
our results are from the continuum regions.
The former is significantly larger 
than the latter; this systematic difference has been seen in 
many other emission line galaxies (\cite{kong99}, \cite{Calzetti97}).

Perelmuter et al. (1995) have obtained spectra for 25 globular cluster
in M81. Following the method of Brodie \& Huchra (1990), based on the 
weighted mean of six indices, they measured these clusters' metallicity.
The mean metallicity was calculated both from the weighted mean of the 
individual metallicities, and directly from the cumulative spectrum of 
the 25 globulars. Both results yielded the same value, 
$\langle{\rm Fe/H}\rangle
=-1.48 \pm 0.19$ $(Z=0.033)$, which is identical to that derived by 
Brodie \& 
Huchra  (1991) using 8 clusters ($\langle{\rm Fe/H}\rangle
=-1.46 \pm 0.31$). No correlation 
has been observed between magnitude and metallicity of the globulars in
the Milky Way and in M31. Thus the mean metallicity of the 25 globulars 
should be representative of the M81 system as a whole. These results 
agree with our results for the metallicity of M81 ($Z \approx 0.03$) very well.
On the other hand,
using the low-dispersion spectra of 10 HII regions in M81, Stauffer \& 
Bothun (1984) estimated the oxygen abundances for those HII regions from 
the observed emission lines. They derived a mean abundance near solar 
and a weak abundance gradient in M81. Using the empirical calibration 
method and the photoionization models, Garnett \& Shields (1987) have 
analyzed the metallicity abundance and abundance gradients for 18 HII 
regions in the galaxy M81. The major result of this study is the 
presence of order-of-magnitude gradient in the oxygen abundance across 
the disk of M81. These results differ from ours. However, the differences
can be readily explained. First, as for the internal reddening, the
previous results are derived from the bright HII regions of M81, 
the abundance gradient is therefore for these
HII regions. But our result come from the whole 
galaxy {\it except} these bright HII regions. So it is not clear that
they should have the same behaviors in metallicity. 
Another caveat to
the previous results is that the Pager et al.
(1980) abundance calibration (which was used in Stauffer \& Bothun 1984, 
Garnett \& Shields 1987) is not expected to be exact. 

At last, we must emphasize that although the method we used in this 
paper can be used to constrain the variation of metallicity, population 
age, and reddening across M81 are for the central region, the bulge, and 
the disk minus the spiral arms, it may not be suited to study the property
of the spiral arms. There are two main reasons. First, there are hundreds 
of HII regions on the spiral arms, and, the evolutionary population 
synthesis methods that we used in this paper are not well represented 
very young clusters. Second, the signal-to-noise ratio decreases from the 
center to the edge of the galaxy, we have smoothed the images with a 
boxcar filter (See \S 2.2). For the outer disk regions, such as the 
spiral arms, the smoothing tends to blend the hundreds of HII regions with 
their surroundings. We will study these bright HII regions at the spiral 
arms of M81 in more detail in a subsequent paper when we have better data, 
using the evolutionary population synthesis method, and show how well it 
can work for young clusters.

\section{Conclusions}

In this paper, we have, for the first time, obtained a two-dimensional SED
of M81 in 13 intermediate colors with the BAO 60/90 cm Schmidt telescope.
Below, we summarize our main conclusions.

\begin{itemize} 

\item Using the new extensive grid of GSSPs covering a wide range of
metallicity and age, we calculated the colors and color indices for
13 colors in BATC intermediate-band filter system. We find that some
of them can be used to break the age and metallicity degeneracy, which
enables us to obtain two-dimensional maps of metallicity, interstellar
reddening and age of M81.

\item From the two dimensional metallicity distribution of M81, we find
no obvious metallicity gradient from the central regions to the outer
disk. In most part of M81, the mean metallicity is about 0.03 with
variation  $\lesssim 0.005$. Some regions in M81, however, have higher
metallicity; they are mostly located in the spiral arms and around HII
regions, where the younger component resides.  

\item From the two dimensional age distribution in M81, we find that the
mean ages of the stellar populations in the central regions are older
than those in the outer regions, which suggests that star formations in
the central regions occurred earlier than the outer regions.

\item We find a strong difference in reddening between the bulge region
and the disk region. In the bulge area, the reddening, E(B-V), is in the
range of 0.08 to 0.15.  The mean reddening in the disk area is higher,
about 0.2. There are some high reddening spiral-like cirrus in the bulge.

\item In order to understand how sensitive our method is to different
assumptions about star formation history and different stellar population
synthesis models, we have studied an exponential star formation history
and compared the results obtained with GSSP and PSSP.  We find that
although the precise values of age, metallicity and interstellar reddening
are different, the general trend of in the metallicity, age and reddening
distributions is similar.

\item  Finally, we have compared the 
internal reddening and metallicity maps of M81 with previous studies.
We find that the agreements are generally good. In addition, we
find that the properties for the bright HII regions and other parts may be
different.

\end{itemize}

The results of M81 presented here illustrate that our method and
observational data provide an efficient way to study the distribution of
metallicity, age and interstellar reddening for nearby face-on galaxies.
Similar data have already been collected for similar galaxies M13,
NGC589 and NGC5055. The analysis results of these galaxies will be
published in a forthcoming paper.

\acknowledgments
We are indebted to Dr. Michele Kaufman for a critical and helpful
referee's report that improved the paper.
We would like to thank A. Bressan, D. Burstein, and G. Worthey for 
useful discussion and suggestion.
We are grateful to the Padova group for providing us with a set of
theoretical isochrones and SSPs. We also thank G. Bruzual and
S. Charlot for sending us their latest calculations of SSPs and
 for explanations of their code. The BATC Survey is supported by the
Chinese Academy of Sciences (CAS), the Chinese National Natural Science
 Foundation (CNNSF) and the Chinese State Committee of Sciences and
Technology (CSCST). Fuzhen Cheng also thanks Chinese National Pandeng
Project for financial support. The project is also supported in part
by the U.S. National Science Foundation (NSF Grant INT-93-01805), and
 by Arizona State University, the University of Arizona and Western
 Connecticut State University.

\end{document}